\documentclass[aps,twocolumn,superscriptaddress]{revtex4}   
\newcommand{\rar}{\ensuremath{\longrightarrow}}
\newcommand{\dzz}{\ensuremath{d_{z^2}}}
\newcommand{\dxxyy}{\ensuremath{d_{x^2-y^2}}}
\newcommand{\dd}{\ensuremath{(d_{z^2}^1 d_{x^2-y^2}^1)}}
\newcommand{\dxy}{\ensuremath{d_{xy}}}
\newcommand{\dxyz}{\ensuremath{d_{\pi}}}
\newcommand{\dxz}{\ensuremath{d_{zx}}}
\newcommand{\dyz}{\ensuremath{d_{yz}}}
\newcommand{\dpi}{\ensuremath{d_{\pi}}}
\newcommand{\pipiest}{\ensuremath{(\underline{\pi},\pi^{*})}}
\newcommand{\piest}{\ensuremath{\pi^{*}}}

\newcommand{\alg}{\ensuremath{a_{1g}}}
\newcommand{\blg}{\ensuremath{b_{1g}}}
\newcommand{\bzg}{\ensuremath{b_{2g}}}
\newcommand{\eg}{\ensuremath{e_{g}}}

\usepackage{amsmath}
\usepackage{graphicx}
\usepackage{epsfig}

\usepackage{color}
\usepackage{bm}

\begin{document}

\title{Ultrafast intersystem crossing in nickel porphyrins}

\author{Javier Fern\'{a}ndez-Rodr\'{\i}guez} 
\affiliation{Department of Physics, Northern Illinois University, DeKalb, IL 60115}
\affiliation{Advanced Photon Source, Argonne National Laboratory, Argonne, IL 60439}

\author{Jun Chang}

\affiliation{College of Physics and Information Technology, Shaanxi Normal University, Xi'an 710062, China}

\author{Arthur J. Fedro}
\author{Michel van Veenendaal}
\affiliation{Department of Physics, Northern Illinois University, DeKalb, IL 60115}
\affiliation{Advanced Photon Source, Argonne National Laboratory, Argonne, IL 60439}



\pacs{82.50.-m  78.47.J-  82.37.Vb  82.53.-k}

\begin{abstract}
We study the relaxation dynamics and intersystem-crossing to the metastable state in laser-pumped tetra and hexa-coordinated nickel porphyrins.
We use a ligand-field model which takes into account the crystal field created by the porphyrin ring and axial ligands.
By accounting for the energy redistribution of the lattice vibrations of the metal-ligand stretch mode
we get an irreversible decay within the order of the hundreds of femtoseconds timescale.
We show how non-equilibrium time-dependent x-ray absorption at
the Ni K-edge measurements can elucidate the nature of the
intermediate states involved in the decay.
Understanding radiationless transitions in this system is of interest for their
relevance in photocatalytic systems and photothermal
sensitizers for cancer treatment.
\end{abstract}

\maketitle

\section{Introduction}

Photoinduced excited-state dynamics is a subject of current
scientific interest.  In pump-probe experiments, an optical pulse is
used to excite a system in the ground state and a second spectroscopy
pulse probes the excited states.  Until recently, techniques like
optical, M\"{o}ssbauer or Raman spectroscopy were used to study the
relaxation process~\cite{Gutlich2004,Kruglik2003}.  The recent
advances in ultrafast pulsed x-ray sources now provide a powerful
method for obtaining knowledge on these excited states
processes~\cite{Hentschel2001,Schoelein2000}.

Metalloporphyrins have been the subject of intensive
research~\cite{Kalyanasundaram1992}.  They have interest as model
systems for biological processes like photosynthesis, electron
transfer, transport of oxygen or photoexcitation and reduction
reactions.~\cite{GoutermanBook,RodriguezKirmaierHolten1989} In the
case of nickel${^\mathrm{(II)}}$ porphyrins the metal center has a
$d^8$ electronic configuration.  In the ground state, the $\dxy$,
$\dxz$ and $\dyz$ orbitals are filled.  The occupation of the $\dzz$
and $\dxxyy$ orbitals depends on the coordination environment of the
Ni metal center.
In a non-coordinating solvent, nickel-porphyrins are four-coordinated
and have a singlet ground state, with the Ni ion in a low-spin
$\dzz^2$ electronic configuration, denoted as $^1$A$_{1g}$ using the
notation for $D_{4h}$ symmetry.  In coordinating solvents, two axial
ligands bind to the Ni metal center and the complex becomes
hexa-coordinated.  The Ni $d^8$ electrons adopt a triplet ground state
configuration $^3\dd$, denoted as $^3B_{1g}$, due to the interaction
with the axial
ligands.~\cite{Kim1983,KimHolten1983,RodriguezHolten1989,RodriguezHolten1990}

With a laser-pump, it is possible to induce a $\pi\rar\piest$
excitation in the porphyrin ring from the porphyrin highest occupied
molecular orbital (HOMO) to the lowest unoccupied molecular orbital
(LUMO) without changing the metal electronic configuration.  The
porphyrin ring excited state decays to a metastable state where the
electronic excitation has been transferred to the metal center.
In 4-coordinated nickel-porphyrins the complex
relaxes in less than 350~fs to a 
metastable $\dd$ state~\cite{RodriguezHolten1989}.  The spin multiplicity of this
state is a matter of debate and it has been assigned as 
both singlet~\cite{BrodardVauthey1999,Kobayashi1979}
and triplet.~\cite{Chirvonyi1981,Kim1983}
In hexa-coordinated nickel-porphyrins
the laser-pumped system undergoes a fast radiationless decay to a
$^1(\dzz^2)$ state.
The double occupation in the $\dzz$ orbital induces the ejection of the
axial ligands and the geometry of the complex becomes square-planar
within a few hundred picoseconds~\cite{RodriguezHolten1990,LinChen2001}.

In this article, we study the relaxation dynamics and intersystem-crossing to the metastable
state in square planar and hexa-coordinated nickel porphyrins.
We use a ligand-field hamiltonian to study the excited states of
nickel-porphyrin and the radiationless decay pathways from the
optically pumped excitation.
By means of a dissipative Schr\"{o}dinger
equation~\cite{JunChang2010,PRLMvV2010,JunChangPRX} we can take into
account the energy redistribution with
the environment and determine the time-scale for
the irreversible relaxation dynamics in this
system.
From the time dependence of the different intermediate state
probabilities we calculate the time-dependent x-ray absorption and
discuss how time depenent x-ray absorption measurements can elucidate
the intermediate states participating in the decay.
Understanding the decay of the photoexcited state is important for
applications, such as the design of photocatalytic systems (fuel
generation by sunlight), or photothermal sensitizers for cancer
treatment, where a fast decay is necessary for converting photons into
vibrational energy and killing tumor cells by
heat.~\cite{Rajapakse2010,Soncin1999,Camerin2005}

\section{Model Hamiltonian}

We consider a ligand-field hamiltonian taking into account
the Ni $d$-shell and the porphyrin ring $\pi$ and $\piest$ orbitals.
We use a full configuration-interaction approach that
takes into account exactly the Coulomb interaction between the 3d
electrons in the metal center.  
For the Slater integrals we use
Hartree-Fock estimates~\cite{CowanBook} scaled to 50\% of their atomic
values to account for the effect of hybridization.
A crystal field of $D_{4h}$ symmetry splits the $d$-orbitals into
different irreducible representations:
$a_{1g} (\dzz)$, $b_{1g} (\dxxyy)$, $e_g$ ($\dpi$, i.e. $\dxz$ and $\dyz$), and $b_{2g} (\dxy)$.
We take for the single-particle energies of th $d$-orbitals similar values to those obtained
from density functional theory (DFT) in
nickel-porphyrins~\cite{LiaoScheiner2006,Patchkovskii2004,ChenRSC2010}.
We take 
$\varepsilon_{\alg}=1.3$, $\varepsilon_{\blg}=3.7$,
$\varepsilon_{\eg}=1.2$, and $\varepsilon_{\bzg}=0$~eV.

In the case of the $\pipiest$ excitation, the energy of the Q-band corresponds in our model
to  the energy difference between the
porphyrin $\pi$ and $\piest$ orbitals:
$\varepsilon_{\piest}-\varepsilon_{\pi}=2.4$~eV for the
four-coordinated complex~\cite{Mizutani1999,KimHolten1983}.  The energy of the Q-band
decreases~\cite{RodriguezHolten1989} to 2.2~eV in the hexa-coordinated
complex.

%

\section{Decay pathway}

The decay is described in terms of two steps of internal conversion that
would recombine the hole in $\pi$ and the electron in $\piest$ with
the metal $d$-shell and a step of intersystem crossing produced by spin-orbit coupling.
The hybridization between the metal center
orbitals and the porphyrin ring orbitals $\pi$ and $\piest$ allowed by
a lowering of the $D_{4h}$ symmetry is responsible for the internal conversion.
The metal center
state  $(\dxyz^3 \dzz^2 \dxxyy^1)$ opens the 
door to intersystem crossing to the
metastable state, since it is directly connected to it via 
spin-orbit coupling (SOC).
%


To calculate the dependence with time of the occupation of the
different quantum states we use an effective
probability-conserving dissipative Schr\"{o}dinger equation
(see refs.~\onlinecite{PRLMvV2010,JunChang2010,JunChangPRX}).    
To obtain an irreversible decay to the metastable state, we include
in our model the exchange of energy of the metal-ligand system with a
thermal bath.
%
%
We approximate the nuclear
wavefunction of the metal-ligand stretching mode as harmonic, and consider the
electron-phonon coupling to be linear, i.e. the vibrational frequencies are
the same for the different electronic states, only the equilibrium
position of the harmonic oscillator changes as the result of an
electronic transition.

The different electronic states couple to each other via spin-orbit coupling
or hybridization.  
For the tetra-coordinated complex, the effective spin-orbit coupling
that we obtain from Hartree-Fock is $\zeta=0.074$~eV (we already take into
account the possibility that there are 3 possible final states).
In the case of the hybridization responsible of internal conversion steps in
the decays we use the value $V_{\pi}=V_{\piest}=0.05$~eV.

For the energy dissipation we consider one totally symmetrical
breathing mode, whose equilibrium distance would depend on the occupation
of the metal center orbitals.
In the square planar complex, we assume the bond-length would be proportional
to the number of electrons in the $xy$ plane (i.e, $\dxy$, $\dxyz$ and $\dxxyy$).
We assume a change in the occupation of one electron would produce
a change in elastic energy of $\epsilon=0.4$~eV in the
square-planar complex and  $\epsilon=0.6$~eV in the hexa-coordinated.
For the frequency of the Ni-N stretching mode we use the value~\cite{XiaoLi1990}
$\hbar\omega=0.05$~eV.  The Huang-Rhys factors would be $g=8$, and $g=12$ phonons
in the tetra and hexa-coordinated cases respectively.
For the observed~\cite{Jia1998,Maclean1996} change in
bond-length of $\Delta Q \approx 0.1$~\AA, these changes in elastic
energy $\epsilon=\frac{1}{2}f\Delta Q^2$ would correspond to a
force constant $f$ of $2.1 \times 10^5$~dyn/cm ($1.3\times 10^3$~eV/nm$^2$) per Ni-N bond.
For the environmental relaxation constant we use the value~$\Gamma^{-1}=40$~fs
similarly to ref.~\onlinecite{JunChangPRX}. 

The obtained time dependence of the probabilities of the four states
involved in the radiationless relaxation of the tetra-coordinated complex is shown in
Fig.~\ref{FIG_DECAY_4COORD}.  
The change in bond-length from the $\pipiest$ excitation and the
dissipation of the energy produced by the Ni-N bond-length
oscillations suppresses the recurrence to the initial $^1\dzz^2\pipiest$ state
in less than 100~fs.  The probability is transferred to the 
intermediate charge-transfer state, 
which is only occupied less than 200~fs, until the excitation
is transferred to the metal center state
$^1(\dxyz^3\dxxyy^2\dzz^1)$, connected by spin-orbit
coupling to the metastable state $^3\dd$.
The system decays to the metastable state within 800~fs.

\section{Time dependent x-ray absorption}

We now turn into the question of how time dependent x-ray absorption
can elucidate which decay path to the metastable state the system is
undergoing.
We calculate the isotropic K-edge quadrupolar x-ray absorption using
Fermi's golden rule.  For this, we exactly diagonalize the ligand-field hamiltonian 
for the initial state and for the final state with a core-hole in 1s.
For the monopolar part of the Coulomb interaction between core and valence we assume
it to be equal to the monopolar part of the valence coulomb interaction, i.e. $U_{1s,3d}=U_{3d,3d}$.
The pre-edge of the calculated spectra is adjusted such that the
absorption features of the metastable state coincide with their position in the experimental
measurements~\cite{ChenRSC2010,March2011}.  We use a core-hole lifetime
lorentzian broadening of $\Gamma=0.7$~eV.
The calculated time evolution of the spectra were convolved with a
50~fs width gaussian to suppress the strong probability oscillations
and account for a finite time resolution.  

\begin{figure}
\begin{center}
\includegraphics[width=0.9\linewidth]{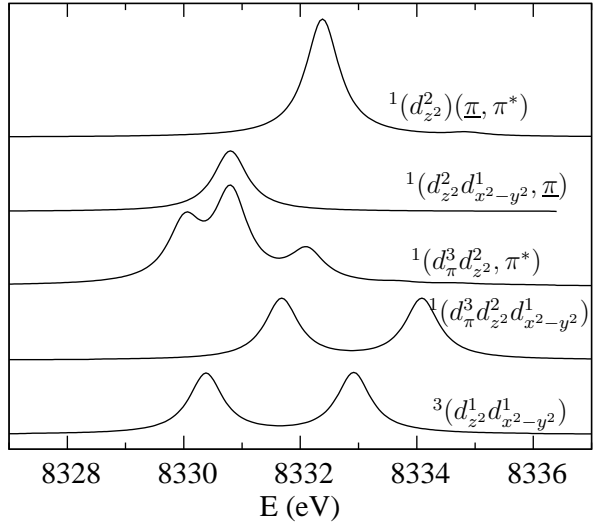}
\end{center}
\caption{\label{indiv4} 
Isotropic K-edge x-ray absorption calculated for the different intermediate states involved in the radiationless decay
of laser-pumped square-planar nickel-porphyrins.
}
\end{figure}

\begin{figure}
\begin{center}
\begin{tabular}{l}
    {\normalsize\textbf{      }} \\ 
    \includegraphics[width=0.7\linewidth]{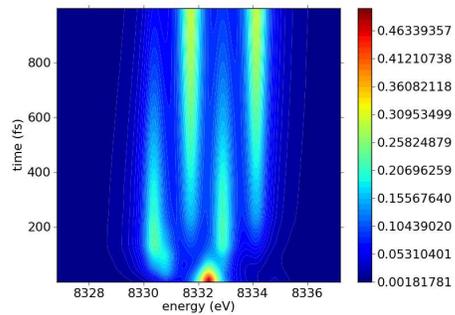}
\end{tabular}
\end{center}
\caption{\label{tXAS4coor} Time-dependent Ni K-edge isotropic x-ray absorption for laser-pumped square-planar nickel-porphyrins.
}
\end{figure}

The calculated spectra for the different intermediate states for the
tetra-coordinated complex are shown in Fig.\ref{indiv4} and the
time evolution of the spectra is shown in Fig.~\ref{tXAS4coor}.
The spectra of the laser-pumped state $^1\dzz^2\pipiest$ corresponds to a single feature 
at 8832.5 produced by transitions to the empty orbital $\dxxyy$.
The metastable state $^3\dd$ and the metal-center intermediate state $^1(\dxyz^1,\dzz^2,\dxxyy^1)$
would both produce a similar absorption spectra 
with two features corresponding to the energies of the singly occupied orbitals.

\section{Conclusions}

We have studied the relaxation dynamics and the possible intersystem-crossing pathways to the
metastable state in photoexcited square-planar and hexa-coordinated nickel-porphyrins.
Our model gives a fast dissipation of the metal-ligand oscillation energy 
and an irreversible decay to the metastable state in a hundreds of femtoseconds timescale.
Radiationless transitions in this system are of interest because of their
relevance for applications, such as the design of catalytic systems,
or photothermal sensitizers for cancer treatment,
where a fast radiationless decay is desirable.

\section{Acknowledgements}

We are thankful to Brian Moritz for useful discussions.
This work was supported by the U. S. Department
of Energy (DOE), Office of Basic Energy Sciences, Division of Materials
Sciences and Engineering under Award No. DE-FG02-03ER46097, the
time-dependent x-ray spectroscopy collaboration as part of the Computational Materials Science Network
(CMSCN) under grant DE-FG02-08ER46540, and NIU Institute for
Nanoscience, Engineering, and Technology. Work at Argonne National
Laboratory was supported by the U. S. DOE, Office of Science, Office of
Basic Energy Sciences, under contract No. DE-AC02-06CH11357.
This research used resources of the National Energy Research
Scientific Computing Center, which is supported by the Office of
Science of the U.S. Department of Energy under Contract
No. DE-AC02-05CH11231.

\bibliography{ManuscriptNiTPP2015}

%


\begin{figure}
\begin{center}
  {\textbf{4-coordinated Ni-P}}
  \includegraphics[width=0.99\linewidth]{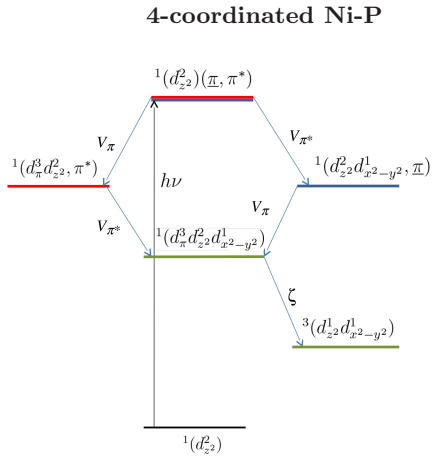}
\end{center}
\caption{\label{FIG4COORD_color}Energy level scheme and possible decay paths in square planar nickel-porphyrins.
The system is initially in the ground state (G.S.) $^1(\dzz^2)$.
The laser pump induces an excitation $\pipiest$ in the porphyrin ring
and the system decays to a metastable state.
The different horizontal postitions of the excited states correspond to the
equilibrium position for each electronic configuration that we assume for the
totally symmetrical breathing mode reponsible for the energy dissipation.  
Possible transitions between states are marked with arrows.
Transitions that change the equilibrium position of the breathing mode
and involve energy dissipation are marked with a one direction arrow,
while those that doesn't change the phononic mode equilibrium position
are marked with a double directional arrow.
%
}
\end{figure}

\end{document}